\documentclass[aps,preprint,superscriptaddress]{revtex4}

\usepackage{graphicx}
\usepackage{float}
\usepackage[small,hang,bf]{caption}
\usepackage{subfigure}
\usepackage{bm}    
\usepackage{amsmath}  
\usepackage{amssymb}
\usepackage{times}
\usepackage{color}

\begin{document}

\title{Partitioning of the molecular density matrix over atoms and bonds} 
\author{Diederik Vanfleteren} 
\affiliation{Ghent University, Center for Molecular Modeling, 
Technologiepark 903, B-9052 Zwijnaarde, Belgium}
\author{Dimitri Van Neck}
\affiliation{Ghent University, Center for Molecular Modeling, 
Technologiepark 903, B-9052 Zwijnaarde, Belgium}
\author{Patrick Bultinck}
\affiliation{Ghent University, Department of Inorganic and Physical Chemistry, Krijgslaan 281 (S3), B-9000 Gent, Belgium}
\author{Paul W. Ayers}
\affiliation{McMaster University, Department of Chemistry, Hamilton, Ontario L8S 4M1, Canada}
\author{Michel Waroquier}
\affiliation{Ghent University, Center for Molecular Modeling, 
Technologiepark 903, B-9052 Zwijnaarde, Belgium}
\date{\today}

\begin{abstract}
A double-index atomic partitioning of the molecular first-order density matrix is proposed. Contributions diagonal in the atomic indices correspond 
to atomic density matrices, whereas off-diagonal contributions carry information about the bonds. The resulting matrices have good localization 
properties, in contrast to single-index atomic partitioning schemes of the molecular density matrix. It is shown that the electron density assigned 
to individual atoms, when derived from the density matrix partitioning, can be made consistent with well-known partitions of the electron density
 over AIM basins, either with sharp or with fuzzy boundaries. The method is applied to a test set of about 50 molecules, representative for various 
types of chemical binding. A close correlation is observed between the trace of the bond matrices and the SEDI (shared electron density index) bond index.
\end{abstract}


\maketitle
\section{Introduction}
The most common way chemists look at molecules is to consider them as composed of atoms held together by chemical bonds. Moreover, the chemical characteristics 
 of atoms and functional groups of bonded atoms are highly transferable between different molecules. Although this picture predates quantum mechanics, it is so useful for 
rationalizing and even predicting experimental observations that it is still ubiquitous. However, the question remains of how to properly 
describe the atom in the molecule (AIM) in a quantum mechanical way. This question has been addressed by many people and various  techniques have been developed 
to describe this elusive concept.

The  techniques used thus far can largely be divided into only a few different categories. In one group, one uses the attachment 
of basis functions to atomic centers to extract the AIM. The best known method of this type is the Mulliken population analysis \cite{mulliken1955}. 
The second, and for the present paper the most important, group of techniques uses a three-dimensional (3D) splitting of space with either sharp boundaries
 between different AIM (e.g.\ Bader's Quantum Chemical Topology (QCT) \cite{bader1994,bader1991,popelier2000}), or with more fuzzy boundaries (e.g.\  the
 original Hirshfeld method \cite{hirshfeld1977} and recent extensions \cite{bultinck20071, bultinck20072, bultinck20073, lillestolen2008, bultinck2009},  
and  Mayer's fuzzy atoms \cite {mayer2004}). In the second group of methods, one uses the molecular electron density and its properties as the guide for  
obtaining the AIM. However, not all AIM properties can be directly expressed in terms of the electron density. A very simple example is the kinetic energy 
of an AIM, for which the full (nondiagonal) density matrix is needed rather than the electron density. For some quantities one even has to go up to 
the second order density matrix. This means that a more fundamental approach to the AIM should be based on density matrices \cite{li1986}. In Bader's QCT only 
the electron density and its derivatives are required to arrive at the AIM energy via the AIM virial theorem \cite{bader1994}, but several concerns remain. 
For instance, Cioslowski and Karwowski arrived at the conclusion that arbitrary choices in the Lagrangian density can have an important influence on the 
uniqueness of Bader's AIM \cite{cioslowski2001}. 

In the present work, we describe how an AIM density matrix that is consistent with a 3D division of the molecular density in AIM's can be obtained. 
Several such methods have been explored previously, for instance, in the work of Alcoba et al \cite{alcoba2005}. These authors derive a QCT-based density 
matrix in the following way. First, the one density matrix is expressed in terms of an orthonormal molecular orbital set through the matrix 
$D =\left\{D_{i\sigma,j\sigma}\right\}$ where $i\sigma$ denotes a molecular orbital $i$ with spin $\sigma$. Given the positive-definite character of 
this matrix, it can be factorized as follows:
\begin{equation}
D_{i\sigma ,j\sigma }^{}  = \sum\limits_{k,l}^{} {\left( {D^{1/2}} \right)_{i\sigma ,k\sigma }  \delta _{k\sigma ,l\sigma } 
\left( {D^{1/2}} \right)_{l\sigma ,j\sigma } }
\end{equation}
The Kronecker delta $\delta _{k\sigma ,l\sigma }$ can be very simply rewritten as:
\begin{equation}
\delta _{k\sigma ,l\sigma }=\left\langle{k\sigma} \vert {l\sigma} \right\rangle=\sum\limits_A {\left\langle {k\sigma} \vert {l\sigma } 
\right\rangle_A}=\sum\limits_A {\left\langle {k\sigma} \lvert {W_A}  \rvert {l\sigma}\right\rangle}
\end{equation}
In other words, the Kronecker delta is written as a sum of atom-condensed overlap integrals $\left\langle {k\sigma} \vert {l\sigma } \right\rangle_A$ where 
the weight $W_A\left({\bm{r}}\right)$ acts as an operator to delineate the AIM domain in the molecule. QCT and Hirshfeld based methods differ mainly in the choice of 
the operator $W_A\left({\bm{r}}\right)$, which is either binary as in QCT, or fuzzy as in Hirshfeld and related methods. An AIM density matrix $D^{A}_{i\sigma ,j\sigma }$ 
with eigenvalues constrained to the interval $\left[0,1\right]$  can then be obtained as:
\begin{equation}
\label{alcoba}
D^{A}_{i\sigma ,j\sigma }  = \sum\limits_{k,l}^{} {\left( {D^{1/2}} \right)_{i\sigma ,k\sigma }  {\left\langle {k\sigma} \lvert {W_A}  \rvert {l\sigma} 
\right\rangle} {\left( {D^{1/2}} \right)_{l\sigma ,j\sigma } }}
\end{equation}
This method, although shown to give interesting results when starting from QCT, has as the drawback of being inconsistent with the underlying AIM method. 
QCT starts from a strict, binary division of space in AIM domains. However, inspection of Eq.~(\ref{alcoba}) shows that the electron density of the AIM is not 
confined to the AIM domain but spreads over the entire space. Extension of the above to a more fuzzy partitioning of space is straightforward, but any scheme along the lines 
 of Eq.~(\ref{alcoba}) will result in orbitals extending far outside the atomic basin assigned to $A$. It should be added that Alcoba et al. also introduced a 
different partitioning by distributing only the molecular occupation numbers in the density matrix over the different atoms, retaining the molecular natural orbitals. 
This obviously again does not lead to very well localized density matrices although the authors reduce this problem by first localizing the molecular natural 
orbitals \cite{alcoba2006,alcoba2007}.\\
In the following we pursue a method that satisfies all the following requirements:
\begin{itemize}
\item The AIM density matrix should be derived starting from a 3D partitioning of space into atomic domains such that the AIM density matrix and electron density always 
remain mutually consistent.
\item The sum of AIM density matrix eigenvalues should equal the electron occupancy of the AIM as obtained from the density analysis, and both the starting AIM density 
and that obtained from the coordinate space diagonal elements of the density matrix should be the same.
\item The density matrices obtained should be localized.
\item The scheme should involve a double atomic index partitioning with diagonal elements AA corresponding to twice the atom A and off-diagonal elements AB, whose 
eigenvectors correspond to chemical bonding between the atoms A and B.
\end{itemize}
The two-index approach is necessary because of the inherent non-local nature of the density matrix, as was recently also argued by Mayer and Salvador \cite{mayer2009}. 
Introducing the two-index partitioning also provides an orbital perspective on the changes in the atoms when bonds are formed, by extracting bond orbitals 
with associated eigenvectors from the ``bond'' density matrices. In this sense, our method is reminiscent in philosophy to the so-called Natural Orbitals for Chemical 
Valence introduced by Nalewajski et al. \cite{nalewajski1994} and used recently to describe chemical bonds by Ziegler and co-workers \cite{michalak20081,michalak20082}.

\section{Theory\label{Theory}}

\subsection{Double atom partitioning of the molecular density matrix}

We use notation $\bm{x}=\bm{r}\sigma$ to specify the single-electron states in coordinate space, where $\sigma$ represents the spin degrees of freedom. The first-order 
density matrix (1DM) for an $N$-electron molecule with  wave function $\Psi (\bm{x}_1 ,\dots ,\bm{x}_N )$ is defined as 
\begin{equation}
\rho (\bm{x},\bm{x}') = N \int d\bm{x}_2 \dots \int d\bm{x}_N 
\Psi^\dagger (\bm{x} ,\bm{x}_2 , \dots ,\bm{x}_N )\Psi (\bm{x}' ,\bm{x}_2 , \dots ,\bm{x}_N )
\end{equation}
We restrict ourselves to molecules with a singlet ground state. In that case 
\begin{equation}
\rho (\bm{x},\bm{x}') = \frac{1}{2}\delta_{\sigma , \sigma'}\rho (\bm{r}, \bm{r}') , 
\end{equation}
and the electron spin can be discarded. 

Successful partitioning schemes based on dividing the molecular electron density $\rho (\bm{r})\equiv \rho (\bm{r},\bm{r})$ in atomic parts rely on introducing
positive atomic weight functions $W_A (\bm{r})$, which set up an atomic decomposition of space \cite{mayer2005} in a sense that $\sum_A W_A (\bm{r})=1$. Then, 
\begin{equation}
\label{AIM}
\rho_A (\bm{r}) = \rho (\bm{r})W_A(\bm{r})
\end{equation}
represents the fraction of the molecular electron density assigned to atom $A$. The various schemes differ mainly in the nature of $W_A (\bm{r})$. In Bader's 
QCT \cite{bader1994,bader1991,popelier2000}, $W_A (\bm{r})$ is a binary operator. In the Hirshfeld method \cite{hirshfeld1977} and the more recent Hirshfeld-I 
extension of it \cite{bultinck20071}, $W_A (\bm{r})$ is given as:
\begin{equation}
W_A (\bm{r})=\frac{\rho^{0}_{A}(\bm{r})}{\sum_B{\rho^{0}_{B}(\bm{r})}}
\end{equation}
where $\rho^{0}_{A}(\bm{r})$ is the density of the isolated atom $A$. The difference between the regular Hirshfeld and the Hirshfeld-I method lies in the choice 
of the states and charges of the atoms used, as described in Bultinck et al. \cite{hirshfeld1977}. In the Iterated Stockholder Atoms 
\cite{lillestolen2008,bultinck2009}, a separate weight function is used for different spherical shells around the atom. 

In this paper we want to extend the idea of dividing the molecular electron density to the (more complex) first-order density matrix. 
At first, a single atom partitioning scheme
\begin{equation}
\rho(\bm{r},\bm{r}') = \sum_A \rho_A (\bm{r}, \bm{r}')
\label{single-index}
\end{equation}
was sought that would fulfill the essential property of hermiticity with eigenvalues between 0 and 2. This single atom density matrix should also be strictly localized; 
that is, the 1DM assigned to an atom $A$ should be only appreciably different from zero when both $\bm{r}$ and $\bm{r'}$ are near atom $A$. Equivalently, the  occupied 
natural orbitals of the atomic density matrix should be strictly confined to the neighborhood of the atom. The transferability of the AIM scheme clearly 
benefits from this concept of localization. Unfortunately, a single atom partitioning scheme of the 1DM as in Eq.~(\ref{single-index}) can never fulfill this 
requirement because the molecular 1DM is (almost always) delocalized and has sizeable contributions for $\bm{r}$ and $\bm{r'}$ near two different atoms.
To accomodate the localization requirement there should be a double atomic weighting, depending on both $\bm{r}$ and $\bm{r'}$. This was also recently argued by 
Mayer and Salvador \cite{mayer2009}. Therefore it seems more natural to introduce a double-atom partitioning. 
\begin{equation}
\rho (\bm{r}, \bm{r}') = \sum_{AB} \rho_{AB} (\bm{r}, \bm{r}'),
\label{eq1}
\end{equation}
where 
\begin{equation}
\rho_{AB} (\bm{r}, \bm{r}') = \frac{1}{2}\left(w_A (\bm{r}) w_B (\bm{r}')+w_B (\bm{r}) w_A (\bm{r}')\right)\rho (\bm{r}, \bm{r}') =\rho_{BA} (\bm{r}, \bm{r}')
\label{deco}
\end{equation}
and the $w_A (\bm{r})$ are atomic weight functions obeying  
\begin{equation}
0\leq w_A (\bm{r})\leq 1 \;\;\;\mbox{and}\;\;\; \sum_A w_A (\bm{r})=1. 
\label{wA}
\end{equation}
We will use the lowercase notation $w_A (\bm{r})$ to indicate weight functions used in the double-index partitioning (\ref{eq1}-\ref{deco}). In general, these will differ from the (uppercase) weight functions $W_A (\bm{r})$ used in a single-index electron density partitioning as in Eq. (\ref{AIM}).

Provided the weight functions are properly localized, it is clear that  $\rho_{AB}(\bm{r} , \bm{r}')$ also will have suitable localization properties, i.e.\ 
being appreciably different from zero only when one of the arguments is near atom $A$ and the other one near atom $B$.

\subsection{Properties of the matrix partitioning}   

The individual contributions to the decomposition (\ref{eq1}) are all hermitian matrices. They clearly come in two types, diagonal ($AA$) and off-diagonal 
($AB$ with $A\neq B$), which will be called atomic density matrices and bond matrices, respectively.  

The diagonal terms $\rho_{AA}$ indeed qualify as first order density matrices, having eigenvalues between 0 and 2. The lower bound is a trivial consequence of 
\begin{equation}  
\rho_{AA}(\bm{r},\bm{r}') = w_A (\bm{r}) \rho(\bm{r},\bm{r}') w_A (\bm{r}') 
\end{equation}
and of the positivity of the molecular 1DM. The upper bound follows from the fact that for any electron wave function $\phi (\bm{r})$ one has 
\begin{eqnarray}
\int d\bm{r}\int d\bm{r}' \phi (\bm{r}) \rho_{AA}(\bm{r} ,\bm{r}') \phi(\bm{r}') 
&=& 
\int d\bm{r}\int d\bm{r}' [w_A (\bm{r})\phi (\bm{r})] \rho (\bm{r} ,\bm{r}')[w_A (\bm{r}')\phi (\bm{r}')]\\ 
&\leq& 
 2 \int d\bm{r} w_A (\bm{r})^2 \phi(\bm{r})^2
\label{ineq1}\\
&\leq& 
 2 \int d\bm{r} \phi(\bm{r})^2 .
\label{ineq2}
\end{eqnarray}
The inequality (\ref{ineq1}) results from the upper bound for the eigenvalues of the molecular 1DM, which implies 
that $\int d\bm{r}\int d\bm{r}' \chi(\bm{r})\chi(\bm{r}')\rho (\bm{r}, \bm{r}') \leq 2 $ for any normalized wave function $\chi (\bm{r})$, and by 
applying this property to $\chi (\bm{r}) = [ w_A (\bm{r})\phi (\bm{r})] /\{\int d\bm{r} [ w_A (\bm{r})\phi (\bm{r})]^2\}^{1/2}$. 
The inequality (\ref{ineq2}) follows from $w_A (\bm{r})\leq 1$. 

The summed atomic density matrices do not carry the total number of electrons, since
\begin{equation}
\sum_A \int d\bm{r} \rho_{AA} (\bm{r} ,\bm{r}) =  \sum_A \int d\bm{r} \rho (\bm{r}) w_A^2 (\bm{r}) 
\leq   \sum_A \int d\bm{r} \rho (\bm{r}) w_A (\bm{r}) \leq N ;  
\end{equation}
here we used Eq.~(\ref{wA}) and $w^2_A (\bm{r})\leq w_A (\bm{r})$. The defect in the electron number must be in the bond matrices, as Eq.~(\ref{eq1}) 
implies that $N=\sum_{AB} \int d\bm{r} \rho_{AB} (\bm{r}, \bm{r})$. For each atom pair $AB$, the bond matrix is seen to carry a positive number of 
electrons, $\int d\bm{r}\rho_{AB}(\bm{r}, \bm{r}) = \int d\bm{r} \rho (\bm{r}) w_A (\bm{r})w_B (\bm{r})\geq 0$. However, the bond matrices are not positive  
by construction, and in general negative eigenvalues do occur. 

The presence of negative eigenvalues in the bond matrices, disturbing at first, can be readily understood by rewriting Eq.~(\ref{deco}) as 
\begin{equation}
\rho_{AB} (\bm{r},\bm{r}') = \sum_i [\psi^{(+)}_{ABi} (\bm{r})\psi^{(+)}_{ABi} (\bm{r}')  -
\psi^{(-)}_{ABi} (\bm{r})\psi^{(-)}_{ABi} (\bm{r}')],
\label{ab}
\end{equation}
where $\psi^{(\pm)}_{ABi}(\bm{r}) = \sqrt{d_i /2} [w_A (\bm{r})\pm w_B (\bm{r})]\psi_i (\bm{r})$, and the $\psi_i (\bm{r})$ and $d_i$ are the natural orbitals and 
corresponding occupancies of the molecular 1DM. The summed atomic density matrices can be rewritten similarly as 
\begin{equation}
\rho_{AA}(\bm{r},\bm{r}')+\rho_{BB}(\bm{r},\bm{r}')= 
\sum_i [\psi^{(+)}_{ABi} (\bm{r})\psi^{(+)}_{ABi} (\bm{r}')  +
\psi^{(-)}_{ABi} (\bm{r})\psi^{(-)}_{ABi} (\bm{r}')],
\label{aa+bb}
\end{equation}

To see what is going on, consider the most naive picture of covalent binding, with a fully occupied ($d_i =2$) molecular orbital 
$\psi_i (\bm{r})\approx [\phi_A (\bm{r}) + \phi_B (\bm{r})]/\sqrt{2}$ as the bonding combination of atomic orbitals $\phi_A (\bm{r})$ and $\phi_B (\bm{r})$. 
Assuming extreme localization properties for 
the weight functions (i.e.\ $w_A (\bm{r})\phi_B (\bm{r})=0$ , $w_A (\bm{r})\phi_A (\bm{r})=\phi_A (\bm{r})$)  one can then interpret the corresponding 
$\psi^{(\pm)}_i (\bm{r}) \approx [\phi_A (\bm{r}) \pm \phi_B (\bm{r})]/\sqrt{2}$ as the bonding/antibonding combination. In this idealized situation, the summed atomic 
density matrices in Eq.~(\ref{aa+bb}) have equal occupancy in the bonding/antibonding orbitals, and the bond matrix in Eq.~(\ref{ab}) needs to have 
negative eigenvalues in order to destroy the occupancy of the antibonding combination and enforce the occupation of the bonding combination in the total molecular 
1DM. Note that negative eigenvalues for $AB$ combinations also occur in the Natural Orbitals for Chemical Valence (NOCV) technique used recen
tly by Ziegler 
and co-workers \cite{michalak20081,michalak20082}.

\subsection{Consistency of density matrix and electron-density partitioning}

It is interesting to reflect on what would be the total electron density $\rho_A (\bm{r})$ assigned to a particular atom $A$ in the double atomic partitioning
 scheme of Eq.~(\ref{eq1}). 
Apart from the diagonal $AA$ density matrix, there are now also contributions from the $AB$ bond matrices, and these can be distributed over the single-atom densities 
in different ways. We will study two of these, as they can be considered to be extreme cases. 

In the nonweighted scheme (this could also be called a Mulliken-like method~\cite{pendas}), the atoms $A$ and $B$ receive an equal share of the density in the bond, 
\begin{equation}
\rho^n_A (\bm{r}) = \rho_{AA} (\bm{r} ,\bm{r}) +\sum_{B(\neq A)} \rho_{AB}(\bm{r} ,\bm{r}) = \rho (\bm{r})w_A (\bm{r}) .
\end{equation}
In the weighted scheme, the atoms receive a share reflecting the balance of the weights $w_A (\bm{r})$ and $w_B (\bm{r})$, 
\begin{eqnarray}
\rho^w_A (\bm{r}) &=& \rho_{AA} (\bm{r} ,\bm{r}) +\sum_{B(\neq A)} \rho_{AB}(\bm{r} ,\bm{r})
\left(\frac{2w_A (\bm{r})}{w_A (\bm{r})+w_B(\bm{r})}\right)\nonumber\\
&=& \rho (\bm{r})w_A (\bm{r}) \left[\sum_B \left(\frac{2w_A (\bm{r})w_B (\bm{r})}{w_A (\bm{r})+w_B(\bm{r})}\right)\right]. 
\label{weightedscheme}
\end{eqnarray}

Summing Eq. (\ref{weightedscheme}) over all atoms yields the molecular electron density, as it should be:
\begin{eqnarray}
\sum_{A} \rho^w_A (\bm{r}) &=& \rho (\bm{r}) \left[\sum_{AB} \left(\frac{2w_A^{2} (\bm{r})w_B (\bm{r})}{w_A (\bm{r})+w_B(\bm{r})}\right)\right] 
\nonumber \\
&=&\rho (\bm{r}) \frac{1}{2} \left[ \sum_{AB} \left(   \frac{2w_A^{2} (\bm{r})w_B (\bm{r})}{w_A (\bm{r})+w_B(\bm{r})}  +  \frac{2w_B^{2} (\bm{r})w_A (\bm{r})}{w_A (\bm{r})+w_B(\bm{r})} \right)\right] 
\nonumber \\
&=&\rho (\bm{r}) \left[ \sum_{AB} \left(   \frac{w_A (\bm{r})w_B (\bm{r}) \left(  w_A (\bm{r}) + w_B (\bm{r})  \right) } {w_A (\bm{r})+w_B(\bm{r})}   \right)\right]    
\nonumber \\
&=&\rho (\bm{r})  \left[ \sum_{AB} w_A (\bm{r})w_B (\bm{r}) \right] = \rho (\bm{r}).
\end{eqnarray}

In both schemes, there is now the possibility of choosing the weights $w_A (\bm{r})$ in such a way that the single-atom density $\rho_A^{n,w}(\bm{r})$ coincides 
with an established AIM electron-density model of the form~(\ref{AIM}). In the nonweighted scheme, one simply takes $w_A (\bm{r})\equiv W_A (\bm{r})$. In the 
weighted scheme, consistency requires that the weights obey a set of nonlinear equations, 
\begin{equation}
\forall A :\;\;\;w_A (\bm{r}) \left[\sum_B \left(\frac{2w_A (\bm{r})w_B (\bm{r})}{w_A (\bm{r})+w_B(\bm{r})}\right)\right]=W_A (\bm{r}).   
\label{nonl}
\end{equation}
In practice, we always found that the iterative sequence 
\begin{equation}
w_A^{(0)}(\bm{r})=W_A (\bm{r});\;\;\;w_A^{(i+1)}(\bm{r})= \left\{\frac{W_A (\bm{r})}
{\left[\sum_B \left(\frac{2 w_B^{(i)} (\bm{r})}{w_A^{(i)} (\bm{r})+w_B^{(i)}(\bm{r})}\right)\right]  }\right\}^{1/2}
\end{equation}
converges rapidly (in at most 20 iterations) to a stable solution of Eq.~(\ref{nonl}) with 
\begin{equation}
\sum_A |w_A^{(i+1)}-w_A^{(i)}| < 10^{-10}
\end{equation}
as the convergence criterion. 

It should be noted that in all numerical work in this paper the Hirshfeld-I weight functions $W_A (\bm{r})$ were taken as input; for other ``fuzzy atom'' prescriptions, like ISA, we expect similar results. The Bader AIM concept is fundamentally different in that it has hard boundaries for the atomic basins. In the absence of non-nuclear
 attractors space is partitioned into the atomic basins, and the corresponding Bader weight function $W_A (\bm{r})$ is zero/one if $\bm{r}$ is outside/inside the basin 
of atom $A$. One can easily verify that for such binary weight functions the nonweighted and weighted schemes coincide, and there is no need to solve the nonlinear
 equations~(\ref{nonl}). On the other hand, the possible presence of non-nuclear attractors is problematic, as it cannot easily be reconciled with the underlying
 AIM picture. However, it deserves to be mentioned that with the present density matrix partitioning, the atomic electron densities from QCT remain localized to the 
AIM basin.

\subsection{Expressions in a finite basis set}

In practice, the partitioned density matrices are expressed in a finite basis set, as used in the molecular calculation. In the basis of the molecular natural orbitals 
$\psi_i (\bm{r})$, e.g., the partitioned 1DM becomes 
\begin{eqnarray}
(\rho_{AB})_{ij}&=& \int d^3r d^3r' \psi_i (\bm{r})\rho_{AB} (\bm{r},\bm{r}')\psi_j(\bm{r}')\nonumber\\
&=& \int d^3r d^3r' \psi_i (\bm{r})\psi_j (\bm{r}')\frac{1}{2}[w_A(\bm{r})w_B(\bm{r'})+w_B(\bm{r})w_A(\bm{r'})]\sum_k d_k \psi_k (\bm{r})\psi_k(\bm{r}')\nonumber\\
&=& \sum_k d_k \frac{1}{2}[C^A_{ik}C^B_{jk}+C^B_{ik}C^A_{jk}] \label{abfinite}.
\label{pab}
\end{eqnarray}
Note that in the present calculation the $\psi_i (\bm{r})$ are just the molecular HF orbitals, and the natural occupations $d_i$ are 2 (0) for the occupied (unoccupied) 
orbitals. It follows that the $\rho_{AB}$ matrix is expressed in terms of the atomic overlap matrix elements (AOM) : 
\begin{eqnarray}
C^A_{ij}& =& \int d^3r \psi_i (\bm{r}) w_A (\bm{r})\psi_j(\bm{r}).
\label{CA}
\end{eqnarray}
The traces of the atomic and bond density matrices correspond exactly to the net and overlap populations for fuzzy atoms as defined in Eq. (6) of Ref. \cite{mayer2004} (of course, when the same weight functions are used). 

The matrix representation $(\rho_{AB})_{ij}$ in Eq. (\ref{pab}), in the finite single-particle space spanned by the basis set, is only equivalent to Eq. (\ref{deco}) in the limit of a complete basis set. But since the underlying SCF calculation is performed in the same finite basis set, it can be argued that consistent calculations actually require the use of the limited basis set expressions in Eq. (\ref{pab}). In a sense, the single-particle space spanned by the finite basis set is "all there is". This holds in particular for subsequent AIM energy considerations: care should be taken that only matrix elements of the electronic Hamiltonian in the finite basis set are used, since these are the ones that fixed the electronic structure of the molecule.

For a reasonable large basis set, the differences between results obtained using Eq. (\ref{pab}) and Eq. (\ref{deco}) are small anyway, as can be suspected by the clean convergence behaviour in Table \ref{tabCOnw}.

\section{Computational Methods}
\label{Computational Methods}

The partitioning scheme described in Sec.~\ref{Theory} was tested by partitioning the 1DM of a small set (listed in Table~\ref{tab1}) of ca. 50 simple molecules with a 
singlet ground state, representative of a diverse variety of chemical bonds. The 1DM was calculated at the Hartree-Fock level of theory using the Aug-cc-pVDZ basis
 set \cite{kendall1992,woon1993,woon2009}. The geometry was taken from a B3LYP \cite{becke1993,lee1988,vosco1980,stephens1994} /cc-pVDZ \cite{dunning1989,woon1993,woon2009} 
optimization.

The scheme was implemented using atomic weight functions $W_A (\bm{r})$ from a Hirshfeld-I analysis. The weight functions $w_A (\bm{r})$ for the double atomic partitioning 
in Eq.~(\ref{eq1}) were constructed in both the nonweighted scheme ($w_A (\bm{r}) = W_A (\bm{r})$) and in the weighted scheme of Eq.~(\ref{nonl}). The iterative Hirshfeld
 weights and the adapted weights $w_{A}(\bm{r})$  are calculated on atom-centered grids, using a logarithmic radial grid of 100 points
 with $r_{\text{min}}=10^{-6}$ \AA\  and $r_\text{max}=20 $ \AA\ , and Lebedev angular grids \cite{lebedev1975,lebedev1976,lebedev1977,lebedev1992,lebedev1995,lebedev1999} 
 with 170 points having randomized orientation on the different shells. 

For CO and C$_3$H$_3$N additional calculations were performed in larger (Aug-cc-pVTZ and Aug-cc-pVQZ )\cite{kendall1992,woon1993,woon2009} basis sets and with larger 
integration grids, in order to assess basis set and grid convergence.

\section{Results and discussion}
\label{Results and discussion}

\subsection{Natural orbitals and populations}
\label{Natural orbitals and populations}
\begin{table}
\centering
\begin{tabular}{|l|l|l|l|l|l|} 
\hline 
AlCl$_{3}$ & C$_{6}$H$_{6}$ (benzene) & CHONH$_{2}$ & H$_{3}$O$^{+}$ & LiF & OS$_{2}$ \\
AlH$_{3}$ & CF$_{4}$ & CHOOH & HCl & LiH & SF$_{6}$ \\
B$_{2}$H$_{6}$ & CH$_{2}$NH & Cl$_{2}$ & HCN & LiOH &  \\
BeH$_{2}$ & CH$_{2}$O & CO & HCOOCH$_{3}$ & N$_{2}$ &  \\
BH$ _{3}$ & CH$_{2}$O$_{2}$ (dioxirane) & CO$_{2}$ & HF & N$_{2}$O &  \\
C$_{2}$H$_{2}$ (acetylene) & C$_{3}$H$_{8}$ (propane) & F$_{2}$ & HNO$_{2}$ & NaCl &  \\
C$_{2}$H$_{4}$ & CH$_{3}$NH$_{2}$ & H$_{2}$ & HNO$_{3}$ & NaOH &  \\
C$_{2}$H$_{6}$ & CH$_{3}$OCH$_{3}$ & H$_{2}$O & HOCl & NH$_{3}$ &  \\
C$_{3}$H$_{3}$N (acrylonitrile) & CH$_{3}$OH & H$_{2}$S & HOOH & O$_{2}$ &  \\
C$_{3}$H$_{4}$ (cyclopropene) & CH$_{4}$ & H$_{2}$SO$_{4}$ & Li$_{2}$ & O$_{3}$ &  \\
\hline 
\end{tabular} 
\caption{\label{tab1}List of molecules in the test set.}
\label{testset}
\end{table}

\begin{figure}[H]
\centering
\includegraphics[width=15cm] {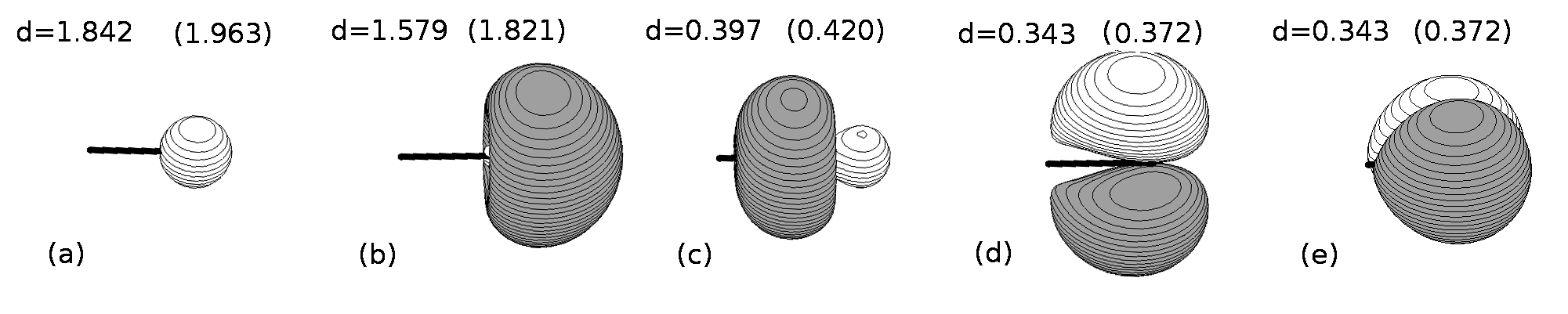}
\caption{The dominant natural orbitals, and the corresponding occupations, of the atomic density matrix of carbon in a CO molecule calculated in the weighted scheme, at the 
HF/Aug-cc-pVDZ level. The occupations in the nonweighted scheme are also indicated, between brackets.     
\label{figCO-CC}}
\end{figure}

\begin{figure}[H]
\includegraphics[width=15cm] {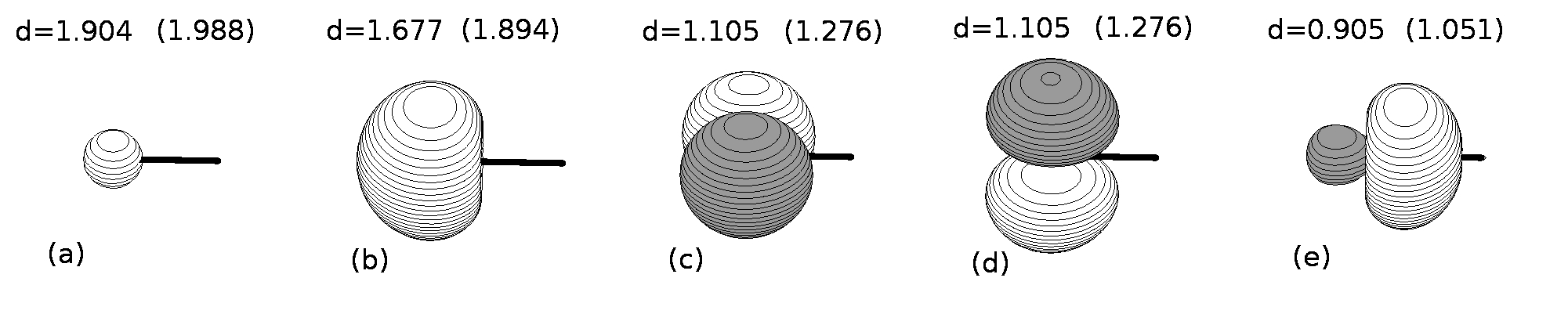}
\caption{The dominant natural orbitals, and the corresponding occupations, of the atomic density matrix of oxygen in a CO molecule calculated in the weighted scheme, at the 
HF/Aug-cc-pVDZ level. The occupations in the nonweighted scheme are also indicated, between brackets.     
\label{figCO-OO}}
\end{figure}

\begin{figure}[H]
\includegraphics[width=15cm] {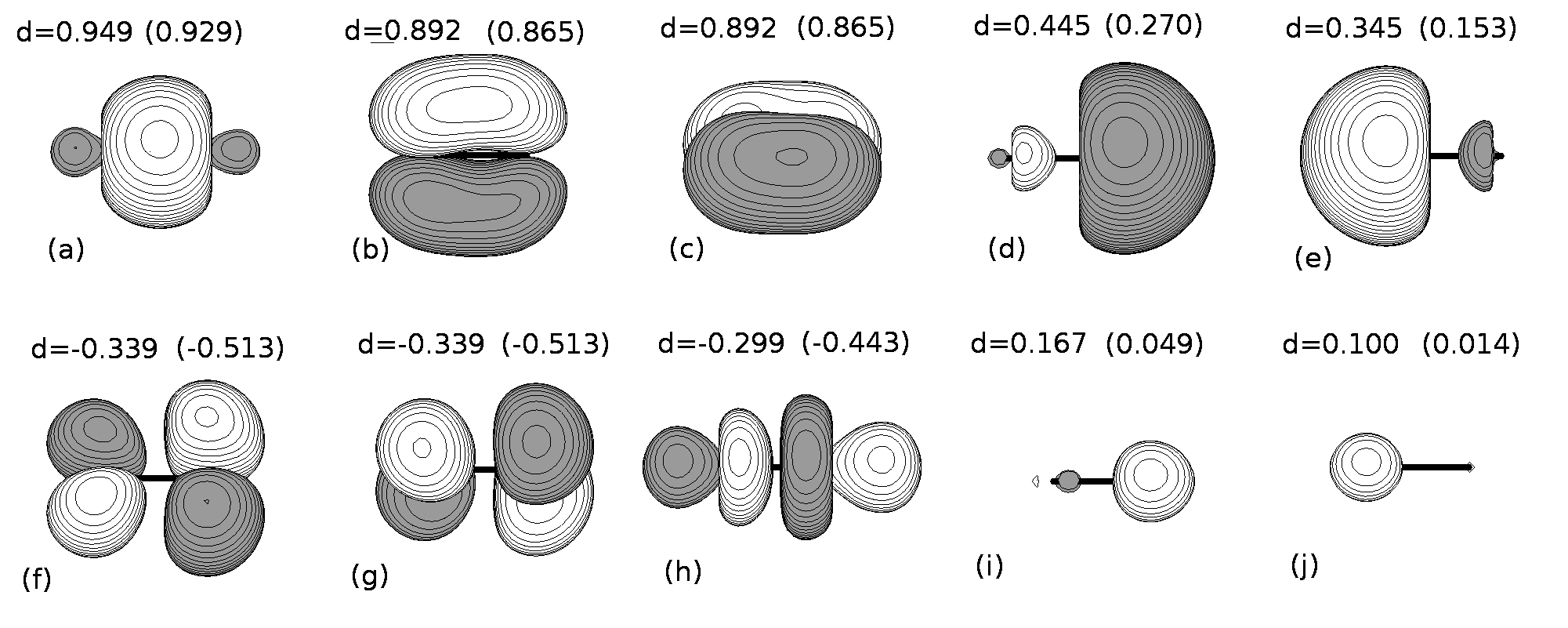}
\caption{The dominant eigenvectors and occupations of (twice) the CO bond matrix in a CO molecule 
calculated in the weighted scheme, at the HF/Aug-cc-pVDZ level. The occupations in the nonweighted scheme are also indicated, between brackets.
\label{figCO-CO} }
\end{figure}

Diagonalisation of the matrices in Eq.~(\ref{abfinite}) leads to the natural orbitals and occupations. As an example of the results obtained, Figs.~\ref{figCO-CC}-\ref{figCO-OO} depict the dominant natural orbitals and occupations of the atomic density matrix $\rho_{AA}$ of carbon and oxygen in CO, within the weighted scheme. The natural orbitals are slightly deformed versions of the typical atomic $1s$ (a), $2s$ (b), $2p_x$, $2p_y$ and $2p_z$ (c-d-e) orbitals (with the $z$-direction corresponding to the internuclear axis). They are clearly localized and have populations between 0 and 2. Note that for these main contributions the shape of the orbitals in the nonweighted scheme is visually indistinguishable from those presented in Figs.~\ref{figCO-CC}-\ref{figCO-OO}. The orbitals that represent the core $1s$ atomic orbitals have a population well below 2.00 (the expected value of an orbital not involved in bonding), resp.\ 1.84 in carbon and 1.90 in oxygen. The rather low eigenvalues for the essentially core orbitals are less pleasing from a chemical point of view. The occupations in the nonweighted scheme are quite different, and are indicated between brackets in Figs.~\ref{figCO-CC}-\ref{figCO-OO}. In the nonweighted scheme, the populations are more in accordance with the full occupation expected of core orbitals, i.e. \ 1.96 and 1.99 for the carbon and oxygen $1s$ orbitals. As discussed later, this difference between both schemes is typical, as the weighted scheme tends to assign a larger fraction of electrons to the bond matrices than the nonweighted scheme. Hence the nonweighted scheme has invariably larger populations for the atomic density matrix natural orbitals. The $2s$ orbitals have  shifted away from the bonding region, to accomodate the free electron pair on C and O  in the CO molecule. However, they are not fully occupied. The $2p_x$, $2p_y$ and $2p_z$ orbitals (not expected to be fully occupied) are shifted to the bonding region.
Figure \ref{figCO-CO} depicts the naturals and occupations of the CO bond matrix $\rho_{AB}$ in the CO molecule. Only the main contributions are shown. Both schemes result in natural orbital shapes that are very similar. There are mainly typical bonding orbitals $\sigma$ (a), $\pi_{1},\pi_{2}$ (b-c)  with positive eigenvalues, and antibonding orbitals $\sigma^*$ (h), $\pi_{1}^*,\pi_{2}^*$ (f-g) with negative eigenvalues. The negative eigenvalues delete antibonding contributions of the $\rho_{AA}+\rho_{BB}$ matrix, whereas the positive eigenvalues reinforce its bonding contributions. The picture also shows two rather ``nonbonding''  
orbitals (d-e) with positive eigenvalues. These serve to reinforce the population of the free electron pairs on carbon and oxygen (the shifted $2s$ orbital on these atoms). 
Note that in the weighted scheme orbitals (i-j) appear which correspond to the $1s$ core orbitals on C and O and have a nonnegligible (0.17 - 0.10) occupation in the bond matrix. 
\begin{table}
\centering
\begin{tabular}{ l l l l l l}
\hline 
    &  C , C        &  O , O          &  C , O           \\  
label  &  eigenvalue   &  eigenvalue     &  eigenvalue      \\
\hline
(a) & 1.842 ( 1.963 ) & 1.904 ( 1.988 ) & 0.474 ( 0.464 ) \\
(b) & 1.579 ( 1.821 ) & 1.677 ( 1.894 ) & 0.446 ( 0.432 ) \\
(c) & 0.397 ( 0.420 ) & 1.105 ( 1.276 ) & 0.446 ( 0.432 ) \\
(d) & 0.343 ( 0.372 ) & 1.105 ( 1.276 ) & 0.223 ( 0.135 )\\
(e)        & 0.343 ( 0.372 ) & 0.905 ( 1.051 ) & 0.173 ( 0.076 ) \\
(f)        & 0.023 ( 0.008 ) & 0.042 ( 0.021 ) & -0.170 ( -0.256 ) \\
(g)        & 0.002 ( 0.000 ) & 0.005 ( 0.001 ) & -0.170 ( -0.256 ) \\
(h)        & $<10^{-13}$   & $<10^{-13}$   & -0.149 ( -0.221 ) \\
(i)        & $\vdots$        & $\vdots$        & 0.084 ( 0.024 ) \\
(j)        &                 &                 & 0.050 ( 0.007 ) \\
$\vdots$ &                 &                 & -0.034 ( -0.061 ) \\
 &                 &                 & -0.007 ( -0.007 ) \\
 &                 &                 & -0.001 ( -0.002 ) \\
 &                 &                 & -0.000 ( -0.000 ) \\
 &                  &                 & $<10^{-13}$\\
 &&&$\vdots$\\
\hline
Sum & 4.528 ( 4.955 ) & 6.743 ( 7.508 ) & 1.364 ( 0.768 ) \\
\hline 
\end{tabular} 
\caption{All natural populations in the atomic density matrices (CC and OO) and bond matrix (CO) in a CO molecule calculated in the weighted scheme, at the HF/Aug-cc-pVDZ level. The occupations in the nonweighted scheme are indicated between brackets. The labels in the first column correspond (for the dominant orbitals) to the labels in Figs.~\ref{figCO-CC}-\ref{figCO-OO}.\label{tab2}} 
\end{table}
Apart from the dominant contributions shown in Figs.~\ref{figCO-CC}-\ref{figCO-CO}, various orbitals with much smaller populations are also present (as the Aug-cc-pVDZ molecular basis set has 46 basis functions). A complete overview is given in Table~\ref{tab2}. Note that for the atomic density matrices, apart from the five dominant natural orbitals, only two more have small ($\sim 10^{-2}-10^{-3}$), while the remainder have vanishing $(< 10^{-13})$ populations. For the bond matrix, apart from the ten dominant orbitals, only four more have small occupations. This can be understood from the fact that the molecule is treated at the HF level, with only 7 (double occupied) spatial orbitals.
One can rewrite the atomic density matrix as
\begin{eqnarray}
\rho_{AA}(\bm{r},\bm{r'})
&=&\sum_{i=1}^{N/2} \left[ w_{A}(\bm{r}) \psi_{i}(\bm{r}) \right] \left[ w_{A}(\bm{r'}) \psi_{i}(\bm{r'}) \right]
\nonumber \\
&=&\sum_{jj'=1}^{N/2} T_{jj'}^{A} \widetilde{\psi_{j}}(\bm{r})\widetilde{\psi_{j'}}(\bm{r'}),
\end{eqnarray}
where $T^{A}$ is the overlapmatrix of the nonorthogonal basis functions $\left[ w_{A}(\bm{r}) \psi_{i}(\bm{r}) \right] $
\begin{eqnarray}
T_{ij}^{A}=\int w_{A}^{2}(\bm{r}) \psi_{i}(\bm{r})\psi_{j}(\bm{r})  \ d^{3}\bm{r}
\end{eqnarray}
and the $\widetilde{\psi_{j}}(\bm{r})$ form a set of orthonormal basis functions 
\begin{eqnarray}
\widetilde{\psi_{j}}(\bm{r})=\sum_{i}(T^{A -\frac{1}{2}})_{ij} \left[ w_{A}(\bm{r})\psi_{i}(\bm{r}) \right]. 
\end{eqnarray}
It is clear that $\rho_{AA}(\bm{r},\bm{r'})$ is a matrix of rank $N/2$, since diagonalisation of the $N/2 \times N/2$ overlapmatrix $T^{A}$ will yield $N/2$ eigenvalues. The same reasoning holds for the bond matrix $\rho_{AB}(\bm{r},\bm{r'})$, but now starting from the 14 (linear independent) functions $\psi_{AB i}^{(\pm)}(\bm{r})$ defined in Eq. (\ref{ab}-\ref{aa+bb}) with $d_{i}=2$ for $i=1,2,...N/2$.

The difference between the nonweighted and weighted scheme is again clear from Table~\ref{tab2}. The differences are sizeable, with consistently smaller populations in the weighted scheme, resulting in traces for the atomic density matrices that are significantly (0.4 - 0.7  ) smaller. This is compensated for by larger eigenvalues in the bond matrix in the weighted scheme, for the orbitals not involved in bonding. The bonding orbitals (a-b-c) in the bond matrix, however, are about equally populated in both schemes.   

\begin{table}[H]
\centering
\begin{tabular}{|l| l| l| l| l| l|}
\hline 
CO & grid=100-170 & grid=500-590 & grid=100-170 &grid=100-170  \\
& Aug-cc-pVDZ & Aug-cc-pVDZ & Aug-cc-pVTZ &Aug-cc-pVQZ  \\
\hline 
A , B & Tr($\rho_{A,B}$) & Tr($\rho_{A,B}$) & Tr($\rho_{A,B}$) &Tr($\rho_{A,B}$)  \\
\hline
C , C   & 4.528 (4.955)& 4.528 (4.955) & 4.529 (4.956)&4.531 (4.958)\\
O , O   & 6.743 (7.508)& 6.743 (7.508) & 6.751 (7.516)&6.752 (7.518)\\
C , O   & 1.364 (0.768)& 1.364 (0.768) & 1.360 (0.764)&1.358 (0.762)\\
\hline 
\end{tabular} 
\caption{Number of electrons present in the atomic density and bond matrices for CO, in the nonweighted and weighted scheme. Bracketed values correspond to the nonweighted scheme. The second column contains the results of the default calculation. In the third column the number of radial and angular grid points is increased to 500 and 590, respectively. In the fourth column a triple-, rather than double-$\zeta$ basis is used, in the fifth colum a quadruple-$\zeta$. \label{tabCOnw}}
\end{table}

In Table~\ref{tabCOnw} the stability of the proposed density matrix partitioning is examined. The results for the matrix traces are clearly converged as far as grid size is concerned, with deviations of less than 0.001 in electron number. The results also seem to be quite stable with respect to basis set size, with differences less than 0.01 going from DZ to TZ, and less than 0.002 going from TZ to QZ.   

\begin{figure}[H]
\centering
\includegraphics[width=5cm] {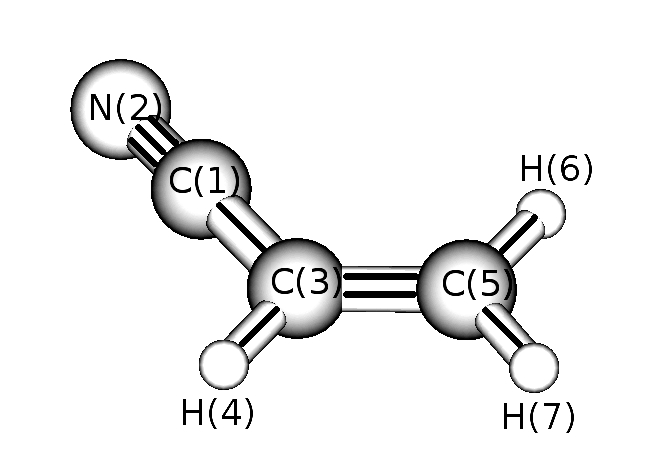}
\caption{Topology of the C$_3$H$_3$N (acrylonitrile) molecule.
\label{figC3H3N:topo}}
\end{figure}

\begin{figure}[H]
\includegraphics[width=15cm] {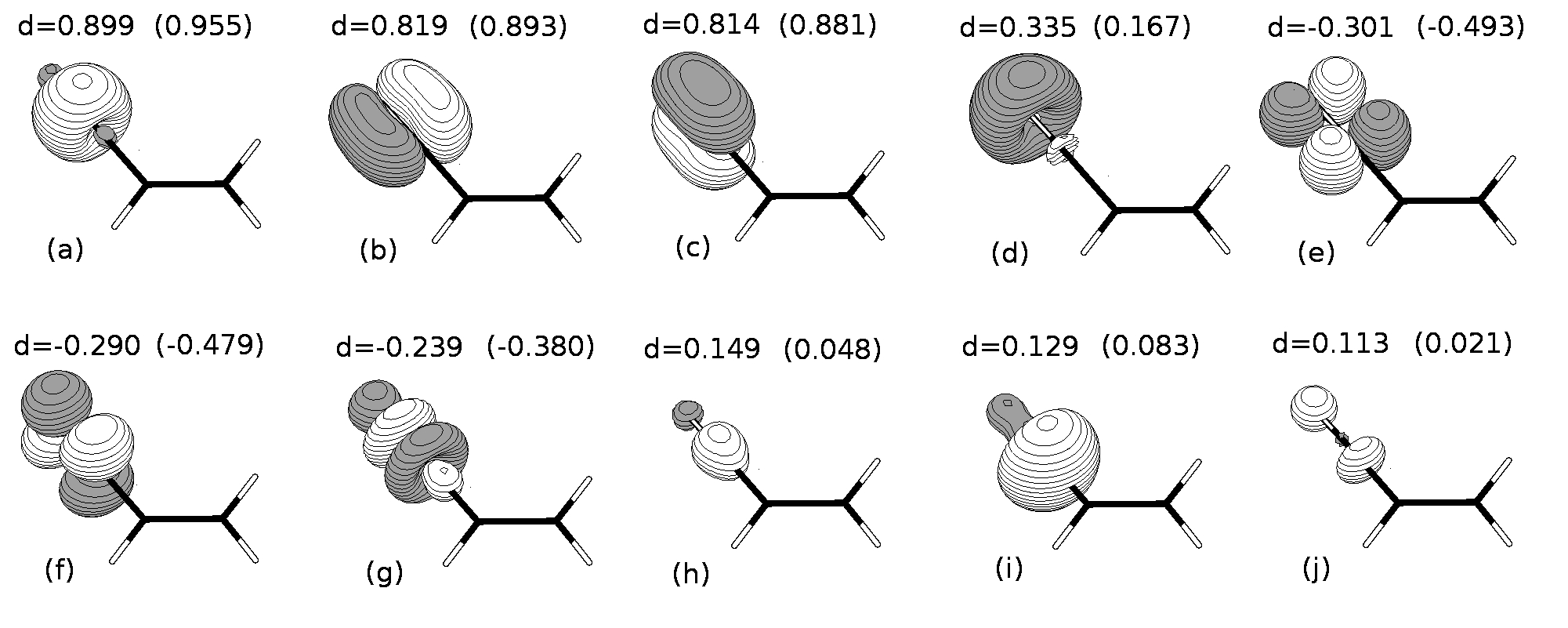}
\caption{The dominant eigenvectors and occupations of the CN bond matrix in a C$_3$H$_3$N molecule calculated in the weighted scheme, at the HF/Aug-cc-pVDZ level. The occupations in the nonweighted scheme are also indicated, between brackets. \label{figC3H3N:CN}}
\end{figure}

As a second example we analyze some results obtained for the density matrix partitioning in $\mbox{C}_3\mbox{H}_3 \mbox{N}$ (acrylonitrile). 
The topology of the molecule is presented in figure \ref{figC3H3N:topo}. Figure \ref{figC3H3N:CN} shows the dominant eigenvectors and occupations of the CN bond matrix. The features are very similar to those of the CO bond matrix. The bond orbitals $\sigma$ (a), $\pi_{1}$ (b)  and $\pi_{2}$ (c) are about equally populated in both schemes. The nonbonding  orbital (d) reinforces the population of the free electron pair on nitrogen. The antibonding 
orbitals  $\pi_{1}^{*}$ (e), $\pi_{2}^{*}$ (f), and $\sigma^{*}$ (g) have negative eigenvalues that delete antibonding contributions of the $\rho_{C(1)C(1)}+\rho_{N(2)N(2)}$ matrix. Small but nonnegligible contributions exist that are complementary to the 1s core electrons on C(1) and N(2) (h and j). All contributions that correct the sum of the atomic density matrices (by deleting antibonding parts and reinforcing nonbonding parts) are significantly larger within the weighted scheme.

\begin{figure}[H]
\centering
\includegraphics[width=15cm] {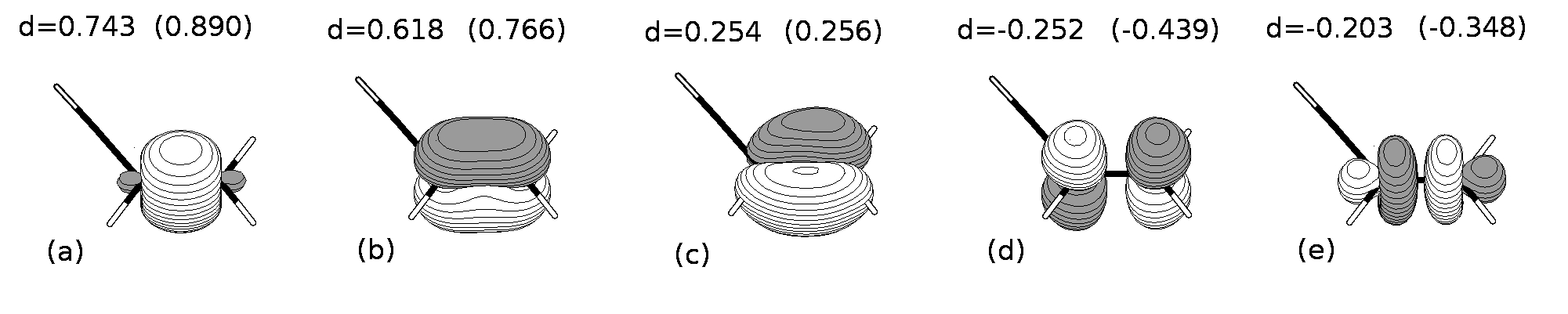}
\caption{The dominant eigenvectors and occupations of the C(3)-C(5) bond matrix in a C$_3$H$_3$N molecule calculated in the weighted scheme, at the HF/Aug-cc-pVDZ level. The occupations in the nonweighted scheme are also indicated, between brackets. \label{figC3H3N:CC2}}
\end{figure}

Figure \ref{figC3H3N:CC2} shows the dominant eigenvectors and occupations of the C(3)-C(5) bond matrix. As it is formally a double bond, large bonding orbitals $\sigma$ (a) and $\pi_{1}$ (b) are expected, next to some smaller antibonding contributions $\sigma^{*}$ (d) and $\pi_{1}^{*}$ (e). There are a lot of small contributions not shown here involving the $2s$ orbitals on both carbons, the $\sigma_{C-H}$ bonds and the $1s$ orbitals on both carbons, that are complementary to the main orbitals of other atomic and bond matrices. However, beyond the many small contributions, there is one considerably larger than the others: a $\pi_{2}$ (c) bond. It's important to mention that this is a rather general feature, also noticed for example in the CH bond matrices of CH$_{4}$, where both $\pi_{1}$ and $\pi_{2}$ contributions are important, although there are no $\pi$ bonds in the CH$_{4}$ molecule. Notice that the trace within the weighted scheme is still larger than the one within the nonweighted scheme, while the bonding orbitals of the 
weighted  scheme have a significantly lower occupation than that of the nonweighted scheme. 

\begin{figure}[H]
\centering
\includegraphics[width=12cm] {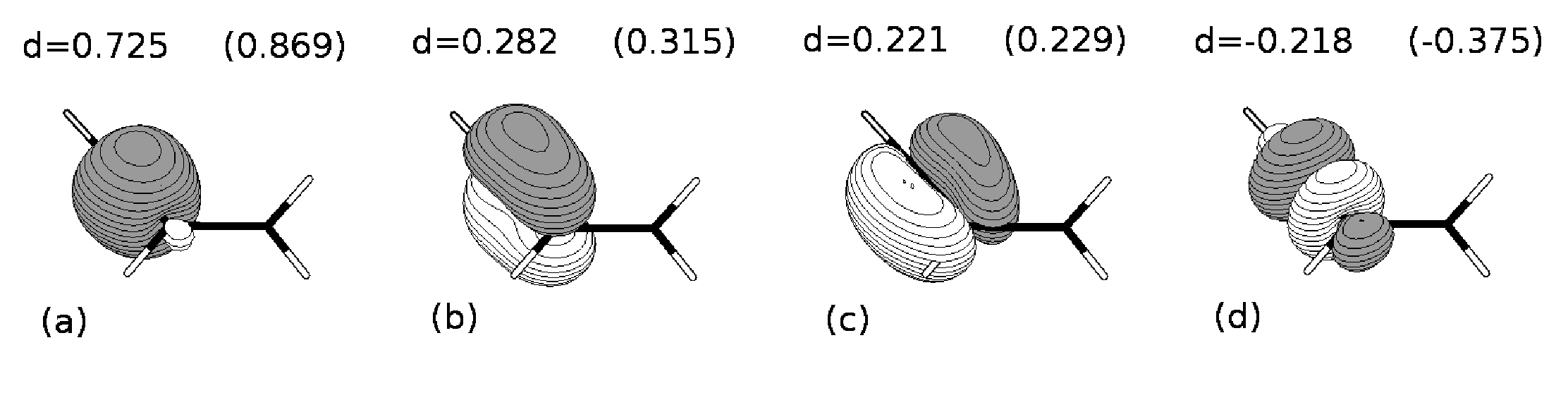}
\caption{The dominant eigenvectors and occupations of the C(1)-C(3) bond matrix 
in a C$_3$H$_3$N molecule calculated in the weighted scheme, at the 
HF/Aug-cc-pVDZ level. The occupations in the nonweighted scheme are also indicated, between brackets.     
\label{figC3H3N:CC1}}
\end{figure}

Figure \ref{figC3H3N:CC1} shows the dominant eigenvectors and occupations of the C(1)-C(3) bond matrix. As it is formally a single bond, a large bonding orbital $\sigma$ (a) is expected, next to a smaller antibonding contribution $\sigma^{*}$ (d). As in the previous case of the C(3)-C(5) double bond, there are a lot of smaller contributions, with two of them predominant: a $\pi_{1}$ (b) and a $\pi_{2}$ (c) orbital. Again, the trace within the weighted scheme is larger than the trace within the nonweighted scheme, while the bonding orbitals have a significantly lower population.

\subsection{Correlation with shared electron density indices}
\label{Shared electrons}
A global test for the partitioning scheme is the evaluation of the total population in its bond matrices. This population should correlate somehow with the bond order. The classical definition of bond order equals it to the electronic occupancy of the bonding orbitals minus that of the antibonding orbitals divided by two. At the Hartree-Fock level of theory, there is a remarkable similarity between these classical bond orders and the results of so-called shared electron density indices (SEDI) \cite{wiberg1968,bader1975,giambiagi1975,mayer1983,fulton1993,angyan1994,mayer2004,fradera1999,ponec2005,mayer2007}. SEDI are obtained from integration of the exchange density at the Hartree-Fock level or exchange-correlation density at correlated levels of theory. The exchange-correlation density reads 
\begin{equation}
\rho^{xc}\left(\bm{r},\bm{r'}\right)=\rho\left(\bm{r}\right)\rho\left(\bm{r'}\right)-\sum_{\sigma\sigma'}\rho^{\left(2\right)}
\left(\bm{r}\sigma,\bm{r'}\sigma' ; \bm{r}\sigma,\bm{r'}\sigma'\right)
\end{equation}
where the diagonal elements of the second-order density matrix $\rho^{\left(2\right)}$ appear. At the single-determinant level, this simplifies to:
\begin{equation}
\rho^{xc}\left(\bm{r},\bm{r'}\right)=\frac{1}{2}\rho\left(\bm{r},\bm{r'}\right)\rho\left(\bm{r'},\bm{r}\right)
\end{equation}
Integrating $\bm{r}$ and $\bm{r'}$ over the atomic domains of atoms $A$ and $B$ and multiplying by two to account for the symmetrical integration over $B$ and $A$ eventually results in:
\begin{equation}
SEDI\left(A,B\right)=4\sum_{i,j}{S^{A}_{ij}S^{B}_{ji}} \\
\end{equation}
where $i$ and $j$ are occupied molecular orbitals, and  $S^{A}_{ij}$  and $S^{B}_{ij}$ are the atom condensed overlaps. 
In the case that the atomic domains are defined by Hirshfeld-I, the atom condensed overlaps reduce to an expression formally very similar to Eq.~(\ref{CA}).
\begin{equation}
S^A_{ij}= \int d^3r \psi_i (\bm{r}) W_A (\bm{r})\psi_j(\bm{r}) 
\label{SA}\\
\end{equation}

The main advantage of SEDI is that these can be computed also between non-covalently bonded atoms, where the classical expression can no longer be used. Given the classical expression for bond order that uses occupancies of bonding and antibonding orbitals, an intriguing question is whether the trace of the bond matrices would also give similar results. In order to answer this question not only for covalently bonded atoms but in general, this section examines the correlation between the trace of bond matrices and the SEDI, both computed starting from the Hirshfeld-I analysis. For the SEDI, the $S^{A}_{ij}$ were constructed using the Hirshfeld-I weights $W_{A}(\bm{r})$ in Eq. (\ref{SA}). The Hirshfeld-I weights were also used to calculate the atomic overlap matrices $C^{A}_{ij}$ of Eq.(\ref{CA}) for the nonweighted scheme. The weights $w_{A}(\bm{r})$ of the nonlinear equations of (\ref{nonl}) were used to build the atomic overlap matrices for the weighted scheme.

\linespread{0.5}
\begin{table} [H]
\centering
\begin{tabular*}{0.61\textwidth}{@{\extracolsep{20 pt}} l l l l }
\hline 
 & $\frac{1}{2}SEDI(A,A)$ & $\mbox{Tr}\;\rho_{AA}^{w}$ & $\mbox{Tr}\;\rho_{AA}^{n}$  \\
\hline
{C $_{ 1 }$ , C $_{ 1 }$} & {3.415} & {3.635} & {4.296} \\
{N $_{ 2 }$ , N $_{ 2 }$} & {5.826} & {5.822} & {6.604} \\
{C $_{ 3 }$ , C $_{ 3 }$} & {3.881} & {3.829} & {4.692} \\
{H $_{ 4 }$ , H $_{ 4 }$} & {0.272} & {0.392} & {0.480} \\
{C $_{ 5 }$ , C $_{ 5 }$} & {4.038} & {3.909} & {4.798} \\
{H $_{ 6 }$ , H $_{ 6 }$} & {0.279} & {0.400} & {0.487} \\
{H $_{ 7 }$ , H $_{ 7 }$} & {0.282} & {0.405} & {0.490} \\
\hline
& $SEDI(A,B)$ & $2 \; \mbox{Tr}\;\rho_{AB}^{w}$ & $2 \; \mbox{Tr}\;\rho_{AB}^{n}$  \\
\hline
\bfseries{C $_{ 1 }$ , N $_{ 2 }$} & \bfseries{2.871} & \bfseries{2.397} & \bfseries{1.615} \\
\bfseries{C $_{ 1 }$ , C $_{ 3 }$} & \bfseries{1.222} & \bfseries{1.294} & \bfseries{0.907} \\
C $_{ 1 }$ , H $_{ 4 }$ & 0.084 & 0.148 & 0.038 \\
C $_{ 1 }$ , C $_{ 5 }$ & 0.197 & 0.261 & 0.076 \\
C $_{ 1 }$ , H $_{ 6 }$ & 0.021 & 0.055 & 0.012 \\
C $_{ 1 }$ , H $_{ 7 }$ & 0.014 & 0.015 & 0.001 \\
N $_{ 2 }$ , C $_{ 3 }$ & 0.253 & 0.319 & 0.065 \\
N $_{ 2 }$ , H $_{ 4 }$ & 0.024 & 0.040 & 0.004 \\
N $_{ 2 }$ , C $_{ 5 }$ & 0.087 & 0.091 & 0.009 \\
N $_{ 2 }$ , H $_{ 6 }$ & 0.010 & 0.024 & 0.003 \\
N $_{ 2 }$ , H $_{ 7 }$ & 0.007 & 0.006 & 0.000 \\
\bfseries{C $_{ 3 }$ , H $_{ 4 }$} & \bfseries{0.914} & \bfseries{0.852} & \bfseries{0.663} \\
\bfseries{C $_{ 3 }$ , C $_{ 5 }$} & \bfseries{1.927} & \bfseries{1.644} & \bfseries{1.228} \\
C $_{ 3 }$ , H $_{ 6 }$ & 0.133 & 0.181 & 0.050 \\
C $_{ 3 }$ , H $_{ 7 }$ & 0.137 & 0.190 & 0.053 \\
H $_{ 4 }$ , C $_{ 5 }$ & 0.132 & 0.182 & 0.050 \\
H $_{ 4 }$ , H $_{ 6 }$ & 0.011 & 0.012 & 0.000 \\
H $_{ 4 }$ , H $_{ 7 }$ & 0.013 & 0.037 & 0.007 \\
\bfseries{C $_{ 5 }$ , H $_{ 6 }$} & \bfseries{0.942} & \bfseries{0.875} & \bfseries{0.673} \\
\bfseries{C $_{ 5 }$ , H $_{ 7 }$} & \bfseries{0.947} & \bfseries{0.883} & \bfseries{0.676} \\
H $_{ 6 }$ , H $_{ 7 }$ & 0.059 & 0.100 & 0.023 \\
\hline 
sum & 28.000 & 28.000 & 28.000 \\
\hline
\end{tabular*} 
\linespread{1.5}
\caption{The SEDI index (second column) versus the traces of the $\rho_{AB}$ matrices in the weighted scheme (third column) and the nonweighted scheme (fourth column) for the C$_3$H$_3$N (acrylonitrile) molecule. AB contributions corresponding to bonded atoms in the Lewis structure are in boldface.}
\label{C3H3N-SEDI}
\end{table}
\linespread{1.5}

Table \ref{C3H3N-SEDI} shows a complete comparison between the SEDI index and the matrix traces within both the weighted and the nonweighted scheme for the C$_3$H$_3$N (acrylonitrile) molecule. It appears that for the triple, double and single bonds present in the molecule (indicated in boldface), the SEDI index is quite close to the classical bond order. The matrix traces of the weighted scheme are usually somewhat lower, while the matrix traces of the nonweighted scheme are much lower (in some cases more than 45 percent). For pairs of atoms that have no classical bond between them, the matrix traces of the weighted scheme are noneglegible and generally somewhat larger than the SEDI indices, whereas the matrix traces of the nonweighted scheme are quite small ($ < 0.08 $). The largest SEDI indices and matrix traces for nonbonded atom pairs can be found in allylic places.

\begin{figure}[H]
\centering
\includegraphics[width=15cm] {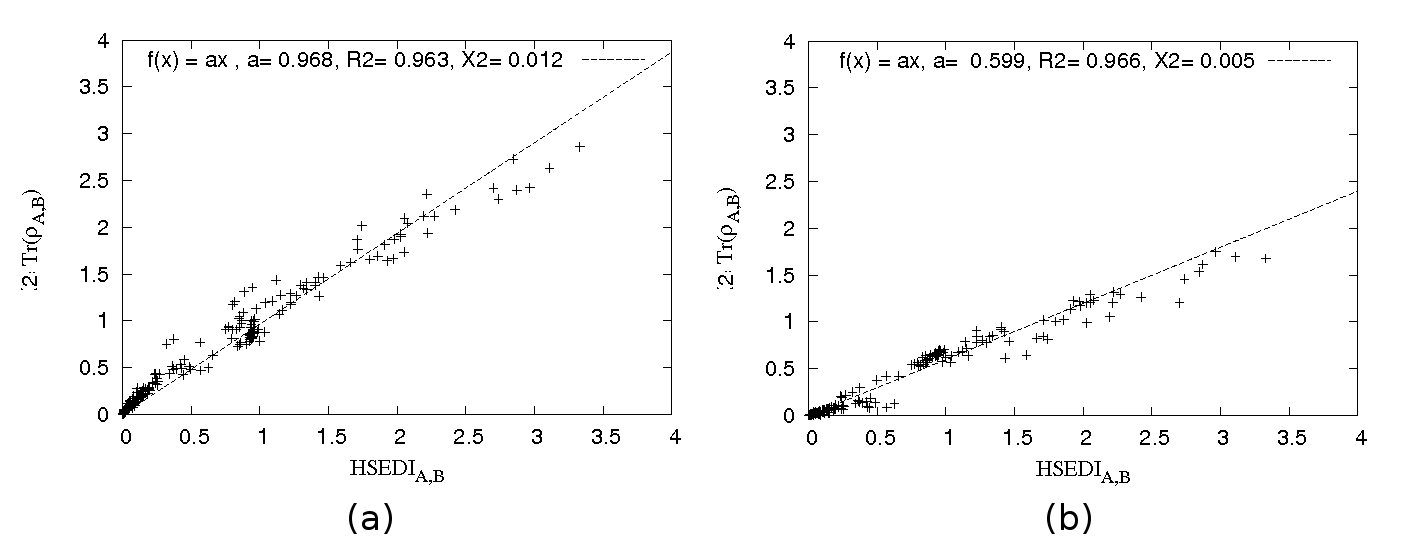}
\caption{Correlation between the SEDI index and the traces of the bond matrices for the weighted scheme (a) and the nonweighted scheme (b)
\label{fig:fullplot-HSEDI-weighted.jpeg}}
\end{figure}

For the test set of table \ref{testset} it was investigated how well the nondiagonal Shared Electron Distribution Index (SEDI) correlates with the matrix traces within both the weighted and nonweighted scheme. A correlation conceptually analogous was first given by I. Mayer for the Mulliken case \cite{mayer1984}. The correlation plots are shown in Fig. \ref{fig:fullplot-HSEDI-weighted.jpeg}. For both schemes there 
is a strong linear correlation ($R^2>0.96$), but the proportionality factor is surprisingly close to unity for the weighted scheme (0.97). By comparison, the slope for the nonweighted scheme is much smaller (0.60). In the case of Hirshfeld-I and the weighted scheme, the SEDI index and the "overlap population" seem to give largely equivalent values $SEDI(A,B) \sim 2Tr(\rho_{AB}^{w})$. In view of the extent (50 molecules) and diversity of the test set, this can hardly be coincidental. One may argue that this is somewhat skewed because of the large number of classically nonbonded atom pairs, for which both measures are close to zero. However, restricting to the bonded atom pairs the proportionality factors hardly change (0.95 for the weighted, 0.59 for the nonweighted scheme). This is all the more surprising, as one would intuitively expect a closer relation between $SEDI(A,B)$ and $2Tr(\rho_{AB}^{n})$ since the weight functions are the same (i.e. $w_{A}(\bm{r}=W_{A}(\bm{r}))$ for the nonweighted scheme. 

The fact that the slope of the correlation plot is larger in the weighted scheme, can be  traced back to the fact that the bond matrices $\rho_{AB}$ 
have a larger trace in the weighted than in the nonweighted scheme, as is evident from Table~\ref{C3H3N-SEDI}. 
The opposite holds for the trace of the atomic density matrices $\rho_{AA}$: these are larger in the nonweighted scheme than in the weighted scheme. Both observations 
can be explained by the Hirshfeld-I weight functions $W_A (\bm{r})$ to be more strongly localized around atom $A$ than the weight function $w_A (\bm{r})$ resulting from the 
nonlinear equations~(\ref{nonl}). This is indeed what is found numerically, and it can also be understood by analyzing Eq.~(\ref{nonl}): in the vicinity of atom $A$, $w_A (\bm{r})$ dominates over the weight functions $w_B(\bm{r})$ of all other atoms. Replacing $w_B(\bm{r})$ in the denominator of Eq.~(\ref{nonl}) by 
$w_A(\bm{r})>w_B(\bm{r})$ therefore leads to 
\begin{equation}
W_A(\bm{r})= \sum_B \frac{2w^2_A(\bm{r})w_B(\bm{r})  }  {w_A(\bm{r})+w_B(\bm{r})}>\sum_B \frac{2w^2_A(\bm{r})w_B(\bm{r})  }  {w_A(\bm{r})+w_A(\bm{r})} =
w_A (\bm{r}) . 
\end{equation}
So the Hirshfeld-I weights $W_A (\bm{r})$  are more localized on the individual atoms and will lead to a smaller number of shared electrons.  

From the preceding discussion it's clear that there seems to be some freedom in choosing an appropriate scheme with corresponding atom weights: one can increase the number of electrons in the overlapmatrix $\rho_{AB}(\bm{r},\bm{r'})$ and get it even close to the SEDI index, at the risk of including some parts (e.g. the core $1s$ electrons) that do not belong there.       
\section{Conclusions}
\label{Conclusions}
%
%


We have introduced a succesful approach to partitioning of a molecular density matrix in constituent atomic and bond contributions. 
The partitioning is such that the density matrices and electron densities for the atoms in the molecule are mutually consistent, and follows from  
requiring that the atom and bond orbitals, as eigenfunctions of the corresponding density matrices, are strictly localized. 
This prevents the use of single atom density matrix partitioning as the molecular density matrix is inherently delocalized.

The atomic density matrices correspond to diagonal elements $\rho_{AA}(\bm{r},\bm{r}’)$, whereas bond matrices correspond to the off-diagonal contributions 
$\rho_{AB}(\bm{r},\bm{r}’)$. Only in the case of the diagonal elements are the eigenvalues restricted to the interval [0,2]. For the bond matrices, 
negative eigenvalues can and do occur.

The weight functions used for partitioning the molecular density matrix were constructed in two different ways: using either directly the weights produced 
from a regular atoms-in-molecules (AIM) density based theory (here Hirshfeld-I), or using a weighted scheme. In both cases the AIM density derived from the 
density matrix is equal to the one obtained directly from the AIM algorithm.

The trace of the bond density matrices was suggested as a useful source of bond indices, loosely related to classical bond orders. A remarkably good correlation 
with shared electron density indices (SEDI) was found, establishing the chemical relevance of the density matrix partitioning. The traces of the bond density matrices 
can in this way be used quite effectively to characterize chemical bonds without requiring second-order density matrices.

It should be mentioned that while the present analysis is restricted to spin singlet molecules, an extension to higher-spin ($S>0$) molecular states seems entirely possible, be applying the decomposition in Eq. (\ref{deco})  to the spin up and down electron density separately,
\begin{eqnarray}
\rho_{AB}^{\sigma}(\bm{r},\bm{r'})=
 \frac{1}{2} \left(w_A (\bm{r}) w_B (\bm{r}')+w_B (\bm{r}) w_A (\bm{r}')\right)\rho^{\sigma}(\bm{r},\bm{r'}) =\rho_{BA}^{\sigma}(\bm{r},\bm{r'}) 
\label{decospin}
\end{eqnarray}
where the atomic weight functions are taken as spin independent.
Certainly for Hirshfeld-like schemes this seems the most natural choice, as the spin-state of the isolated atoms building up the weight functions is washed away when these are considered as molecular constituents. Note that even for the simpler problem of molecular electron density partitioning, spin considerations have hardly been studied and should be explored further.  

\section{Acknowledgements}

We acknowledge support from Dr. Toon Verstraelen (CMM) who provided us with a Hirshfeld-I program and 
from Dieter Ghillemijn (Phd, Ghent University) for comparative calculations. DVF  acknowledges support from the research council (BOF) 
of Ghent University and FWO-Vlaanderen. PWA acknowledges support from Sharcnet, NSERC, and the Canada Research Chairs. 
%
%
   \bibliographystyle{unsrt}

\end{document}